# Temperature-invariant magneto-optical Kerr effect in noncollinear antiferromagnet

Camron Farhang,[1,+] Weihang Lu,[1,+] Yuchuan Yao,[2] Pratap Pal,[2] Shaofeng Han,[1] Jian-Guo Zheng,[3] Hua Chen,[4,5] Chang-Beom Eom,[2] and Jing Xia[1,*]

[1] *Department of Physics and Astronomy, University of California, Irvine, Irvine, CA 92697, USA*
[2] *Department of Materials Science and Engineering, University of Wisconsin-Madison, WI 53706, USA.*
[3] *Irvine Materials Research Institute, University of California, Irvine, Irvine, CA 92697, USA*
[4] *Department of Physics, Colorado State University, Fort Collins, CO 80523, USA*
[5] *School of Materials Science and Engineering, Colorado State University, Fort Collins, CO 80523, USA*

Noncollinear antiferromagnets exhibit anomalous Hall and magneto-optical Kerr effects driven by Berry curvature despite negligible net magnetization, promising ultrafast spintronic applications. While both effects are theoretically expected to reveal the intrinsic Berry curvature that serves as a spintronic memory bit, their quantitative interpretation is complicated by additional temperature-dependent contributions superimposed on the magnetic order parameter: extrinsic skew scattering in dc Hall transport, and optical-resonance effects in visible-wavelength Kerr measurements. Here we perform polar Kerr measurements at the infrared telecommunication wavelength (1550 nm) on epitaxial, stoichiometric $Mn_3NiN$ single crystal films, revealing for the first time a spontaneous Kerr signal that remains stable within a few percent over a 200 K range below the Néel temperature. This temperature-invariant intrinsic Kerr response contrasts with the strongly temperature-dependent anomalous Hall effect in the same sample dominated by extrinsic skew scattering. Our findings establish infrared Kerr effect as a robust, local probe of Berry curvature in noncollinear antiferromagnets, enabling quantitative characterization and advancing antiferromagnetic spintronic technologies.

The Hall effect [1] and its optical counterpart, the magneto-optic Kerr effect (MOKE) [2], are direct consequences of time-reversal symmetry breaking (TRSB) in magnetic systems. In paramagnets, both effects scale with the applied magnetic field [1,2] and are widely used in magnetic field sensors [3,4]. In ferromagnets, which exhibit spontaneous magnetization without a magnetic field, the zero-field anomalous Hall effect (AHE) and MOKE scale with the magnetization [5,6] and are commonly used for its readout. Driven by the demand for faster information processing, there has been growing interest in TRSB systems with nominally zero net magnetization ($m_{net}$), such as antiferromagnets (AFM) [7,8], altermagnets [9,10], and quantum spin liquids [11]. It has been shown theoretically [7] and experimentally [12] that non-collinear antiferromagnets can host spin textures with spontaneous TRSB, and exhibit unexpectedly large AHE [12] and MOKE signals [8,13] despite negligible $m_{net}$. Importantly, regardless of FM or AFM, the intrinsic contribution to the nonzero AHE can be connected to the Berry curvature $\Omega(k)$ in momentum space [7], a geometric property of Bloch electrons:

$$\sigma_{xy} = \frac{-e^2}{2\pi^3\hbar}\int_{BZ} dk\, \Omega(k) \qquad (1)$$

, where $\sigma_{xy}$ is the anomalous Hall conductivity (AHC), $\Omega(k) = \sum_n f[\epsilon_n(k) - \mu]\, \Omega_n(k)$ is the total Berry curvature over the occupied bands in the Brillouin zone (BZ) ($f$ is the Fermi distribution function; $\epsilon_n(k)$ is the band energy; $\mu$ is the chemical potential).

Based on equation (1), the experimentally measured $\sigma_{xy}$ should, in principle, provide a direct and quantitative measure of the Berry curvature integral $\int_{BZ} dk\, \Omega(k)$ associated with the spin texture, the effective "information carrier" in AFM spintronic devices. However, this ideal scenario is not realized in practice. Although the Berry curvature integral should be essentially temperature independent deep in the AFM phase, once the AFM order parameter has saturated, the measured AHC $\sigma_{xy}$ exhibits a strong temperature dependence ($Mn_3Sn$ [12], $Mn_3Ge$ [14], $Mn_3Pd$ [15], $Mn_3NiN$ [16]), often varying by an order of magnitude. This temperature dependence in transport, as already well known in the studies of FM conductors [17] and reexamined recently for AFM materials, [15] is ultimately due to the intrinsic contributions being intertwined with extrinsic scattering mechanisms such as skew scattering, which are difficult to evaluate and remove.

When a non-collinear antiferromagnet is irradiated by light at an optical frequency $\omega$, the electrons oscillate over diminishing distances scaling as $1/\omega^2$. As a result, the optical Hall conductivity $\sigma_{xy}(\omega)$ (the transverse conductivity induced by the oscillating electric field at optical frequencies) becomes insensitive to extrinsic scattering mechanisms [18]. If the photon energy is kept low enough to avoid optical transitions, the polar MOKE signal $\theta_K(\omega)$ (equation 2) [8] could serve as a quantitative probe of an ac version of the Berry curvature.

$$\theta_K(\omega) = \mathrm{Re}\left[\frac{-\sigma_{xy}(\omega)}{\sigma_{xx}(\omega)\sqrt{1+i\left(\frac{4\pi}{\omega}\right)\sigma_{xx}(\omega)}}\right] \qquad (2)$$

To date, however, temperature-dependent MOKE studies [19,20] on non-collinear antiferromagnets, all performed at relatively high photon energies above $1.5\ eV$, reflect both the genuine temperature dependence of the magnetic order parameter in those materials and optical-resonance contributions, including even sign reversals [20]. Isolating the intrinsic, order-parameter-saturated response by Kerr measurements alone has therefore remained challenging for quantitative readouts of AFM spintronics based on non-collinear antiferromagnets.

Here we perform MOKE measurements at a much lower optical energy of $0.8\ eV$, corresponding to the widely used $1550\ nm$ telecommunication wavelength. Using high-quality single crystal $Mn_3NiN$ (Fig.1a) epitaxial films as a model non-collinear antiferromagnet we demonstrate for the first time a temperature-invariant MOKE response when the AFM order is fully established at low temperatures, which is in sharp contrast to the strongly temperature-dependent AHE that is dominated by skew scattering.

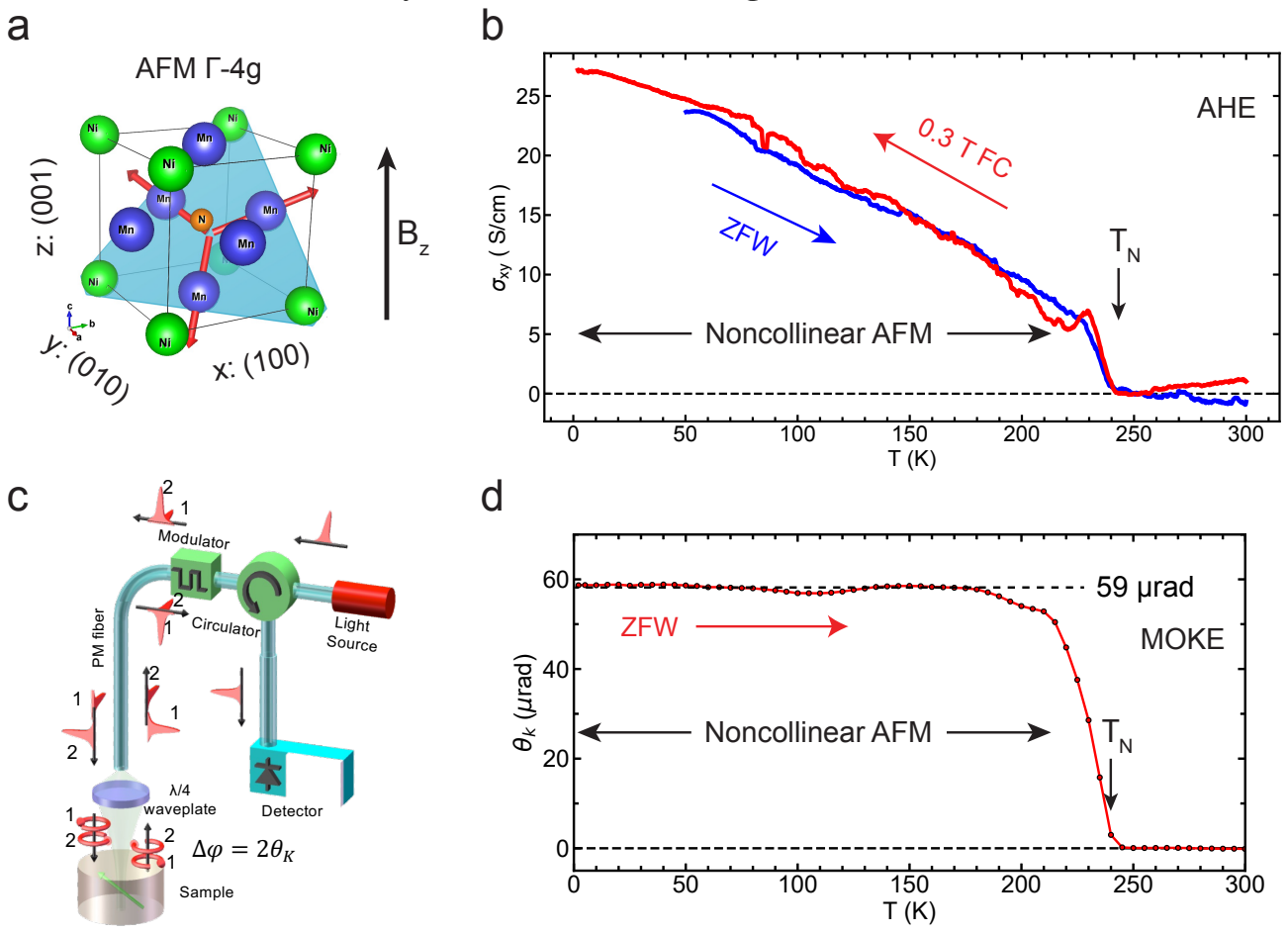

**Fig. 1. Noncollinear AFM $Mn_3NiN$ with strongly temperature-dependent AHE and temperature-invariant MOKE. a,** $\Gamma_{4g}$ spin texture in $Mn_3NiN$. **b,** AHE during *0.3 T* field cool (FC) and subsequent zero-field warm (ZFW) showing strong temperature dependence. **c,** Schematics of a zero-area-loop fiber-optic interferometer microscope for polar MOKE $\theta_K$ measurements at $0.8\ eV$ photon energy. **d,** $\theta_K$ measured during ZFW showing temperature-invariance below $200\ K$.

The $40\ nm$ thick $(001)$ $Mn_3NiN$ films were grown on LSAT substrates via sputtering. Their detailed structural, transport, and magnetic characterizations have been reported elsewhere [21]. Below the Néel temperature $T_N{\sim}240\ K$, it adopts a so-called $\Gamma_{4g}$ AFM spin texture [21,22] as illustrated in Fig.1a, with Mn moments (red arrow) forming a triangular pattern in the (111) kagome plane (blue). Neutron diffraction measurement of a similar $Mn_3NiN$ film has been reported in Ref [21] and is replotted in the supplemental Fig. S1. The magnetic (001) neutron diffraction intensity emerges sharply below $T_N \sim 240\ K$, and remains nearly constant down to $2\ K$, indicating a temperature invariant AFM order parameter. A small net moment $m_{net}$ of $\sim 0.008\ \mu_B/Mn$, due to spin-orbit coupling (SOC) and TRSB, enables magnetic field ($B$) switching of the chirality of the $\Gamma_{4g}$ AFM spin texture through Zeeman coupling. This chirality switching is accompanied with sign reversals of the Berry curvature $\Omega(k)$, net moment $m_{net}$, AHE signal $\sigma_{xy}$, and MOKE signal $\theta_K$.

MOKE measurements are performed using a zero-loop Sagnac interferometer microscope [23–26] in the polar geometry (Fig.1c) operating at the $1550\ nm$ telecommunication wavelength, below the epsilon-near-zero (ENZ) point [27] to suppress optical transitions. This technique offers exceptional sensitivity, routinely achieving 10 nrad resolution, which is critical for testing temperature invariance. Its superior precision stems from its unique design that exclusively detects microscopic time-reversal symmetry breaking (TRSB) while rejecting non-TRSB effects such as anisotropy and vibrations. Unlike traditional MOKE methods, the Sagnac approach does not require hysteresis measurements to eliminate non-TRSB contributions, which is important for $Mn_3NiN$, whose switching field exceeds $9\ T$ below $175\ K$ [21].

Because of this prohibitively large switching field, AHE measurements were performed using a special method to remove the large longitudinal resistivity contributions, as detailed in the supplemental information. As shown in Fig.1b, the AHE $\sigma_{xy}$ turns on sharply at $T_N$, reaching $5\ S/cm$ just a few Kelvins below $T_N$. Upon further cooling, $\sigma_{xy}$ continues to rise almost linearly to $27\ S/cm$ at $2\ K$. As we discuss later, this linear increase in $\sigma_{xy}$ is likely due to extrinsic skew scattering.

In contrast, the spontaneous MOKE signal measured during zero-field warming (ZFW) is fully saturated at $\theta_K = 59.0 \pm 0.2\ \mu rad$ below $200\ K$, deep in the AFM state (Fig.1d). This demonstrates that unlike AHE, MOKE at $1550\ nm$ is a robust probe of intrinsic Berry curvature.

This temperature-invariance of the spontaneous MOKE is observed in full magnetic hysteresis, and is largely uniform across the sample. Fig.2a shows $\pm 9\ T$ hysteresis loops measured at a $2\ \mu m$ spot with both optical beam and the magnetic field aligned along the (001) direction. The applied magnetic field selects one of the two $\Gamma_{4g}$ AFM spin texture chiralities via Zeeman coupling (or 4 out of 8 if considering all equivalent [111] orientations of the Mn kagome planes), which remains stable after field removal. The field dependence of AHE and MOKE on top of hysteresis is likely to be due to the field-induced canting of Mn moments. At $210\ K$, just below the Néel temperature, the hysteresis loop (Fig.2a, red) shows a remnant (spontaneous) MOKE signal $\theta_K(B = 0) \sim 50\ \mu rad$ and a saturation value $\theta_K(B = 9\ T) \sim 190\ \mu rad$. A spatial scan (Fig.2b) reveals a uniform signal with $\pm 3\ \mu rad$ variation, well above the Sagnac interferometer's noise floor of $0.01\ \mu rad$, indicating some intrinsic inhomogeneity. At $175\ K$ (Fig.2a, blue), the $9\ T$ field is barely sufficient to fully polarize the chiral domains, yielding a slightly reduced remnant signal of $46\ \mu rad$. At $20\ K$, full hysteresis cannot be achieved with our maximum field of $9\ T$, resulting in a non-hysteretic S-shape (Fig.2a, green). The corresponding zero-field MOKE image (Fig.2c) shows nearly zero spontaneous Kerr signal, consistent with alternating chiral domains smaller than the beam size cancelling each other.

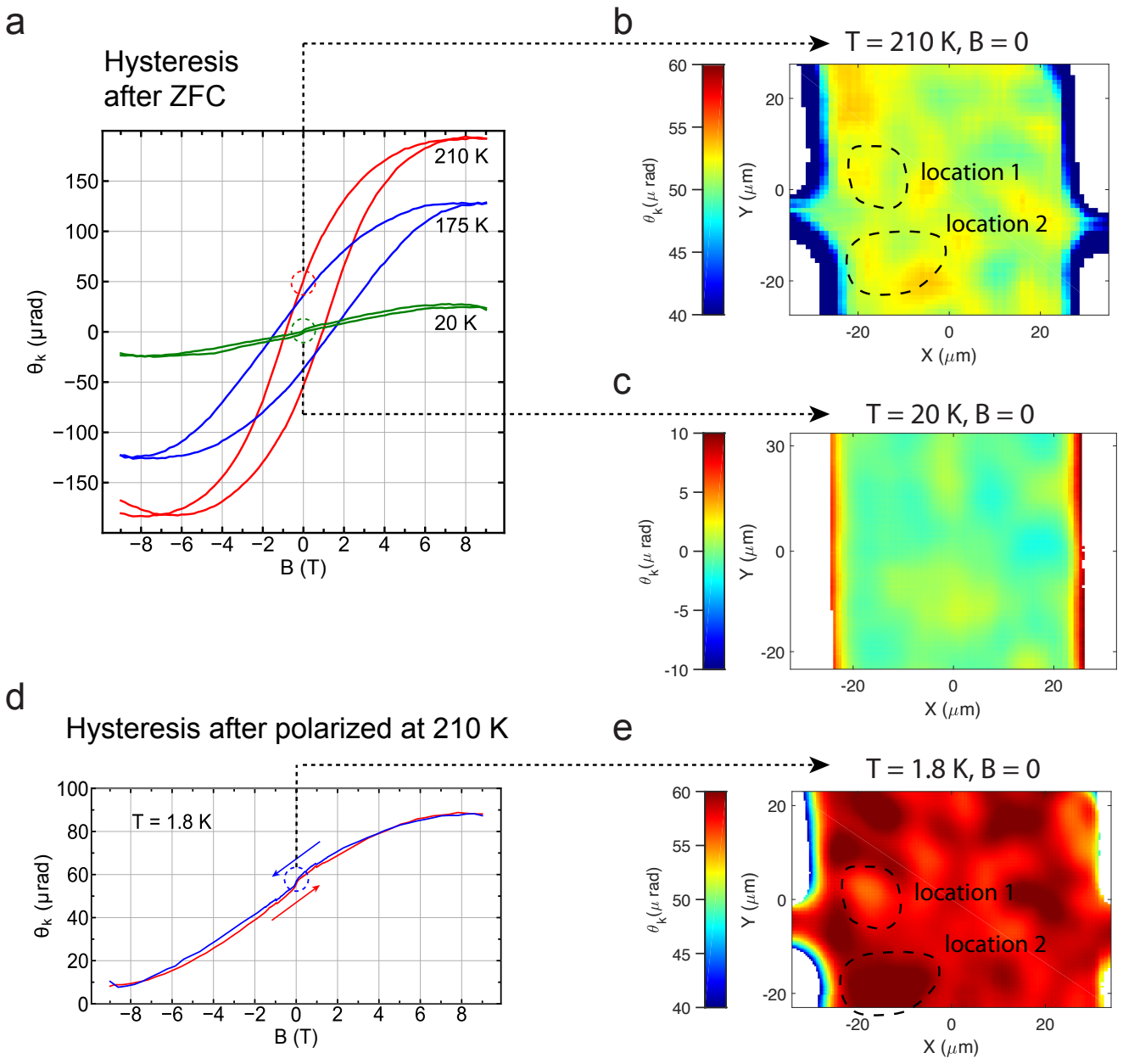

**Fig. 2. MOKE $\theta_K$ and its spatial imaging during hysteresis. a,** MOKE hysteresis at various temperatures after zero field cool (ZFC). Below 175 K, a 9 T magnetic field is not sufficient to polarize the Berry curvature domains during hysteresis. **b, c,** Zero-field (ZF) MOKE imaging at 210 and 20 K respectively. **d,** MOKE hysteresis at 1.8 K after the Berry domains have been polarized during a hysteresis at 210 K and are subsequently cooled in ZF to 1.8 K. Abnormal jumps in $\theta_K$ are visible near ZF in hysteresis below 50 K. **e,** ZF MOKE imaging at 1.8 K of the same region to the 210 K image as shown in **b**, when the Berry domains were polarized during hysteresis.

The divergence of the switching magnetic field at low temperatures, also found in similar samples in Ref. [21], highlights the potential of $Mn_3NiN$ films for memory applications resistant to external perturbations. Chirality "training" can be performed slightly below $T_N$ using a moderate magnetic field before further cooling. Fig.2d shows the hysteresis at 1.8 $K$ after such training for 210 $K$. The remnant MOKE signal at 1.8 $K$ is 54 $\mu rad$, nearly identical to the 50 $\mu rad$ value measured at 210 $K$, which supports the temperature-invariance Berry-curvature of a robust $\Gamma_{4g}$ noncollinear magnetic order. The absence of hysteresis further confirms that at low temperatures, a moderate field of 9 $T$ is insufficient to switch domain chirality once it was set at higher temperature. Interestingly, a sudden 3 $\mu rad$ jump occurs in the low-field region near $\pm 0.1$ $T$ as highlighted by the blue dashed circle in Fig.2d. A similar jump is also seen at 20 $K$ in Fig.2a as highlighted by the green dashed circle. The origin of this feature is unclear and warrants future investigation.

The corresponding MOKE image at 1.8 $K$ (Fig.2e) shows a uniform signal with a similar $\pm 3$ $\mu rad$ spatial variation seen at 210 $K$ (Fig.2b). This reproducible, above-noise-floor spatial variation indicates a genuine real-space inhomogeneity of the Berry-curvature response superimposed on the temperature-invariant signal. A full characterization of this real-space texture and its evolution is reported in a separate study [28].

We next evaluate the precision of the temperature invariance of the spontaneous MOKE. ZFW measurements after specific field-cooling (Figs. 3a) were performed at two locations (locations 1 and 2 in Figs. 2b and 2e), which showed signals of 54 and 59 $\mu rad$, respectively, at 1.8 $K$ (Figs. 2e). Prior to ZFW, the sample was either fully polarized at 210 $K$ using a 9 $T$ field or cooled through $T_N = 240K$ under a 0.3 $T$ field. As shown in Fig. 3a, both methods yield indistinguishable MOKE signals, indicating equivalent domain polarization. Below 100 $K$, the MOKE signals remain remarkably stable, varying by less than 1% (Figs. 3b and 3d), and between 100 $K$ and 210 $K$ the variation stays within a few percent. The temperature-invariant signal and the magnitude of the small spatial variation are reproducible across thermal cycling; the detailed spatial pattern of the variation evolves between cycles and is analyzed separately [28].

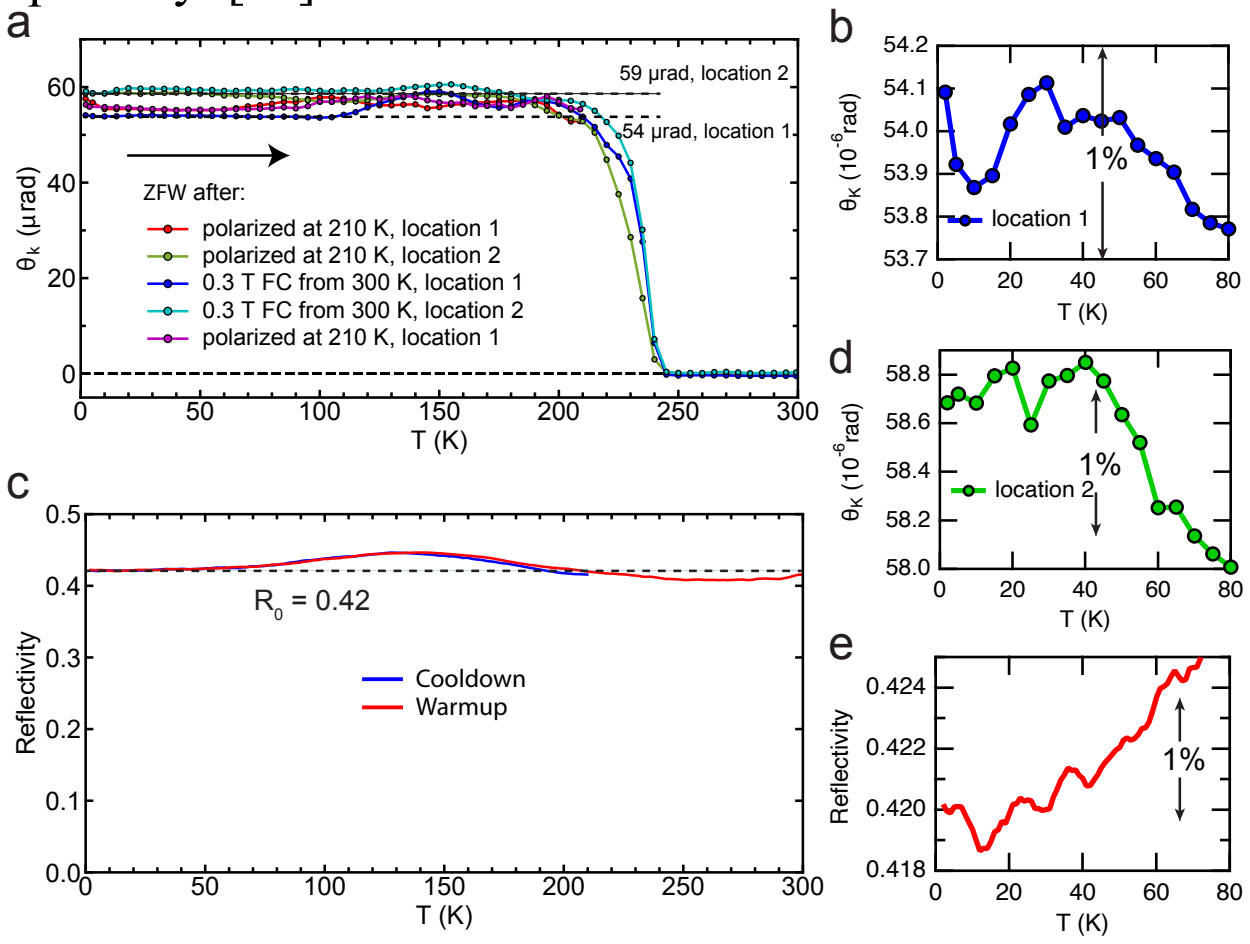

**Fig. 3. Temperature-invariant MOKE and reflectivity. a, b, c,** MOKE signal $\theta_K$ during zero-field warm (ZFW) at two locations, either after 0.3 T FC through $T_N$ or after the chirality domains are polarized during hysteresis at 210 K. **c, e,** Optical reflectivity R with very weak temperature dependence.

MOKE depends on both $\sigma_{xy}(\omega)$ and $\sigma_{xx}(\omega)$ via equation (2), promoting us to examining the temperature dependence of $\sigma_{xx}(\omega)$. Since $\sigma_{xx}(\omega)$ is linked to the dielectric function $\varepsilon(\omega)$ as $\sigma_{xx}(\omega) = -i\varepsilon_0\omega[\varepsilon(\omega) - 1]$, and $\varepsilon(\omega)$ is related to optical reflectivity $R(\omega)$ by $R(\omega) = \left|\frac{\sqrt{\varepsilon(\omega)}-1}{\sqrt{\varepsilon(\omega)}+1}\right|^2$, we could infer the temperature dependence of $\sigma_{xx}(\omega)$ by measuring temperature dependence of reflectivity. As shown in Fig. 3c, the reflectivity $R$ at 0.8 $eV$ photon energy is exceptionally stable from 1.8 $K$ to 300 $K$. Below 80 $K$, $R(0.8\ eV)$ varies by less than 1% (Fig. 3e), indicating a nearly temperature-independent $\sigma_{xx}(0.8\,eV)$. According to equation (2), this supports a temperature-invariant $\sigma_{xy}(0.8\,eV)$. The dielectric function of Mn₃NiN films have recently [27] been measured at room temperature in the $0.5-6\,eV$ range, with a reported $\varepsilon(0.8\,eV)\sim 1.6+37i$ and $\sigma_{xx}(0.8\,eV)\sim(670-10i)S/cm$ for Mn₃NiN/LSAT. With the measured $\theta_K(0.8\,eV)=59\,\mu rad$, we can use equation (2) to estimate the temperature-invariant $\sigma_{xy}(0.8\,eV)\sim 0.3\,S/cm$.

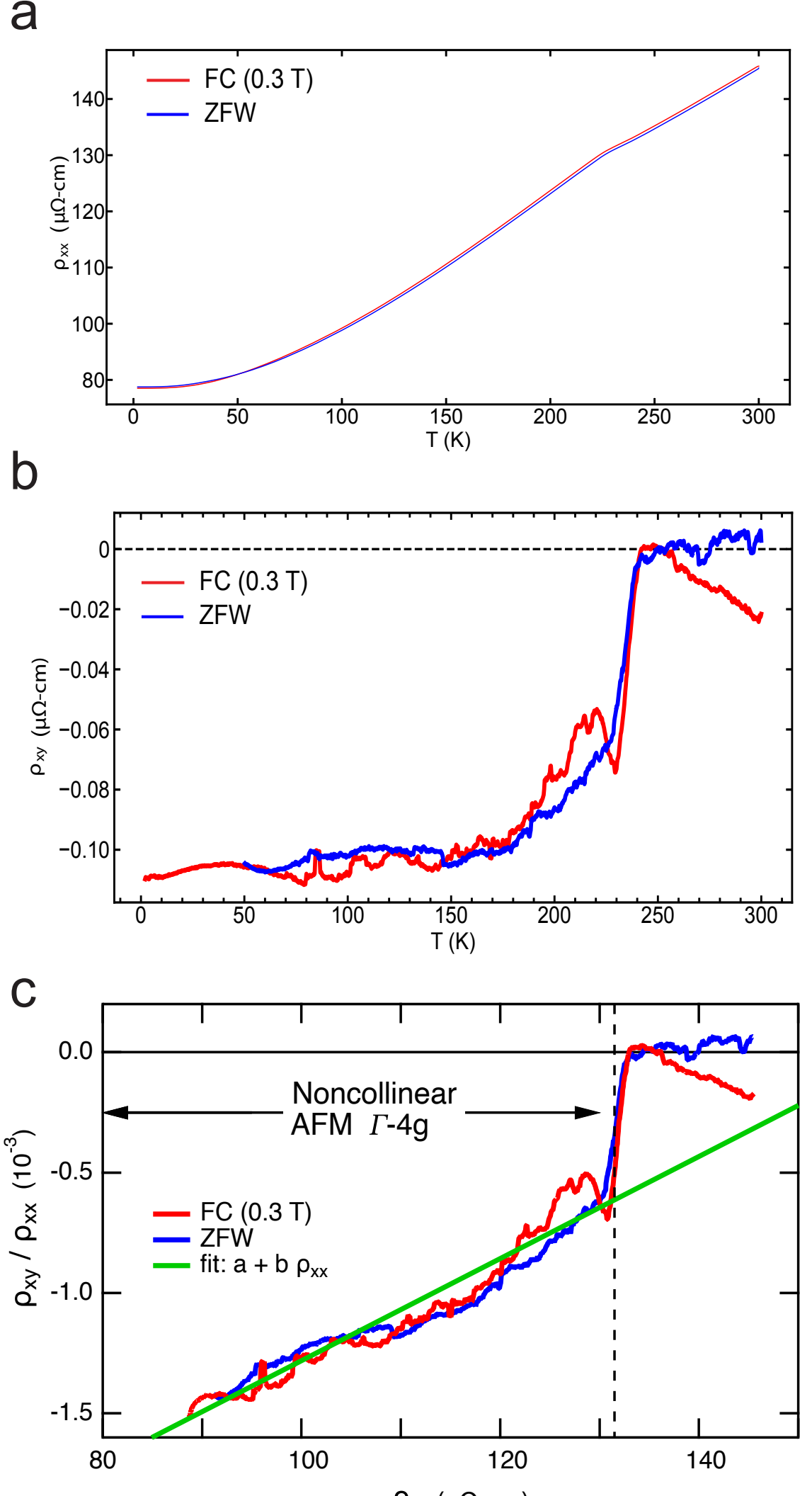

**Fig. 4. AHE and its scaling law. a,** $\rho_{xx}$ measured during $\pm 0.3\,T$ FC. **b,** $\rho_{xy}$ during $0.3\,T$ field cool (FC) (red) and subsequent zero-field warm (ZFW) (blue). **c,** scaling of $\rho_{xy}/\rho_{xx}$ vs. $\rho_{xx}$ with a a linear fit (green) in the AFM phase, suggesting contributions from both intrinsic Berry curvature and extrinsic skewing scattering.

To understand the pronounced and puzzling temperature-dependence of DC Hall conductivity $\sigma_{xy}$ (Fig. 1b), we apply the standard scaling analysis motivated by a relaxation time approximation [17,15]. Since the skew scattering contribution scales with the transport relaxation time $\tau$ and is therefore inversely proportional to $\rho_{xx}$ [29], while the intrinsic contribution from Berry curvature and side-jump scattering is independent of $\tau$, the Hall resistivity is expected to following the empirical scaling relation:

$$\rho_{xy} = a\,\rho_{xx} + b\,\rho_{xx}^2 \qquad (3)$$

, where $a$ represents skew scattering and $b$ corresponds to the intrinsic contribution. As shown in Fig. 4a, $\rho_{xx}$ decreases by half during cooling due to reduced phonon scattering, in stark contrast to the temperature invariant optical conductivity $\sigma_{xx}(0.8\,eV)$. Conventional measurement of $\rho_{xy}$ is challenging below $175\,K$ as full hysteresis can no longer be achieved with a $9\,T$ field. Instead, we perform $\pm 0.3\,T$ field cooling (FC) from above $T_N$ to train the chirality, followed by ZFW. Subtracting the raw Hall data (Supplementary Fig. S3a) removes the longitudinal $\rho_{xx}$ contribution, which is even under time-reversal. The resulting $\rho_{xy}$ curves during $0.3\,T$ FC and the subsequent ZFW (Fig. 4b) show near-perfect overlap, confirming that a *0.3 T* field is too weak to alter the spin texture. Fig. 4c plots $\rho_{xy}/\rho_{xx}$ versus $\rho_{xx}$, along with a linear fit (green) within the AFM phase. The intercept and slope correspond to the parameters $a=-3\times10^{-3}$ representing the Hall angle due to skew scattering, and $b=25\,S/cm$ due to the intrinsic contribution in equation (3). The excellent agreement with the linear fit strongly indicates that a significant extrinsic skew scattering contribution makes a major contribution to the pronounced temperature dependence of $\sigma_{xy}$. In contrast, the optical Hall conductivity $\sigma_{xy}$ probed by MOKE is insensitive to the scattering time and reflects predominantly the intrinsic Berry-curvature contribution; because the dc and optical responses sample $\sigma_{xy}$ at different frequencies, their magnitudes need not coincide, and it is their contrasting temperature dependence, which is pronounced in dc transport, while absent in the optical response, that isolates the intrinsic component.

We further note that the observed temperature-invariant MOKE depends on stoichiometry. As described in the Supplementary Information, we also grew nitrogen-deficient Mn₃NiN thin films under a low nitrogen partial pressure, with a slightly higher Néel temperature of 255 *K*. The spontaneous MOKE signal during ZFW is presented in Fig. S2. Although the peak Kerr signal $\theta_K(241\,K)=59\,\mu rad$ is comparable to that of the stoichiometric film, $\theta_K$ decreases significantly at lower temperatures. We speculate that below 200 *K*, a mixed $\Gamma_{5g}$ phase emerges in nitrogen-deficient films, stabilized by local lattice distortions of the Mn₆N octahedra [30] caused by nitrogen deficiency [31]. Because the symmetry of $\Gamma_{5g}$ phase forbids AHE [16,22,32] and therefore MOKE, the Kerr signal is suppressed until the temperature rises above ~230 *K*, where the $\Gamma_{4g}$ phase once again dominates.

At optical frequencies, electrons oscillate over extremely short distances, making them largely insensitive to scattering from defects and phonons. This makes optical probes particularly promising in revealing intrinsic quantum geometrical properties of electronic systems. In addition to Berry curvature, other geometrical quantities that depend on derivatives of Bloch states with respect to crystal momentum, such as quantum metric [33–36], Berry curvature dipole [33,37], quantum connection [33], etc., have played important roles in the modern understanding of nonlinear [33,37–39] optical and transport responses. At the linear response level, our work demonstrates that MOKE at low photon energies such as $0.8\,eV$ serves as an effective, direct probe for Berry curvature that would otherwise be difficult to isolate by transport means. At frequencies above the Drude peak, in a fixed-band picture, the intrinsic $\sigma_{xy}(\omega)$ and inter-band $\sigma_{xx}(\omega)$ that determine $\theta_K$ only weakly depend on temperature through the Fermi-Dirac distribution function. $\theta_K$ is therefore expected to be much less temperature sensitive than dc conductivities, in which scattering plays a dominant role. If the photon energy gets close or exceeds the plasma frequency, however, pronounced resonances and many-body effects start to prevail, and temperature-induced quasiparticle energy shifts and broadening begin to strongly affect the optical conductivity. This, together with the genuine temperature dependence of the magnetic order parameter in those materials, likely underlies the previously reported temperature-dependent MOKE in non-collinear AFM [19,20], all conducted above $1.5\,eV$. The choice of $0.8\,eV$ $(1550\,nm)$ in this work is thus one practical choice within the spectral window where $\sigma_{xx}(\omega)$ is temperature-stable and below optical transitions: it lies below the plasma frequency of materials like $Mn_3Sn$ $(1.2\,eV)$ [40], and coincides with the most widely used telecommunication wavelength. Optical components and ellipsometers at $1550\,nm$ are readily available, making this wavelength ideal for characterizing Berry curvature-driven magneto-optics and advancing ultrafast AFM spintronics.

This project was supported by the Gordon and Betty Moore Foundation EPiQS Initiative, Grant # GBMF10276 and NSF award DMR-2419425 awarded to J.X.. C.B.E. acknowledges support for this research through a Vannevar Bush Faculty Fellowship (ONR N00014-20-1-2844) and the Gordon and Betty Moore Foundation's EPiQS Initiative, Grant GBMF9065 awarded to C.B.E.. H.C. acknowledges support by NSF grant DMR-1945023 and DMR-2531960 awarded to H.C.. J.X. conceived and supervised the project. C.F., W.L., S.H., and J.X. carried out the experimental measurements. Y.Y., P.P., and C.B.E. fabricated and characterized the samples. H.C. provided theoretical support. J.G. Zheng performed Energy-dispersive X-ray spectroscopy (EDX) characterization. J.X. analyzed the data and drafted the paper with the input from all authors. All authors contributed to the discussion of the manuscript.

Additional experimental details are provided in the Supplemental Material [41], which includes Refs [42,43].

+*These authors contributed equally.*
**Email: xia.jing@uci.edu*

**Data availability**
Source data have been deposited in a figshare repository (*10.6084/m9.figshare.30508892*).

## Supplementary Information for

# Temperature-invariant magneto-optical Kerr effect in noncollinear antiferromagnet

Camron Farhang,[1,+] Weihang Lu,[1,+] Yuchuan Yao,[2] Pratap Pal,[2] Shaofeng Han,[1] Jian-Guo Zheng,[3] Hua Chen,[4,5] Chang-Beom Eom,[2] and Jing Xia[1,*]

*[1] Department of Physics and Astronomy, University of California, Irvine, Irvine, CA 92697, USA*

*[2] Department of Materials Science and Engineering, University of Wisconsin-Madison, WI 53706, USA.*

*[3] Irvine Materials Research Institute, University of California, Irvine, Irvine, CA 92697, USA*

*[4] Department of Physics, Colorado State University, Fort Collins, CO 80523, USA*

*[5] School of Materials Science and Engineering, Colorado State University, Fort Collins, CO 80523, USA*

*[+] These authors contributed equally.*

**Email: xia.jing@uci.edu*

**Contents:**

**Magnetic neutron diffraction:**

The antiferromagnetic order parameter is proportional to the magnetic neutron diffraction intensity, which can be measured in thicker films with larger sample volumes. Neutron diffraction measurements of stoichiometric 250 nm $Mn_3NiN$ films, grown under conditions similar to the 40 nm film discussed in the main text, were reported in Ref. [1]. In Fig. S1, we replot those neutron data alongside the spontaneous magneto-optic Kerr effect (MOKE) signal from the 40 nm film. The $Mn_3NiN$ (001) magnetic diffraction peak emerges below the *Neel* temperature, and its intensity remains nearly constant down to 2 *K*, consistent with the temperature-invariant MOKE signal.

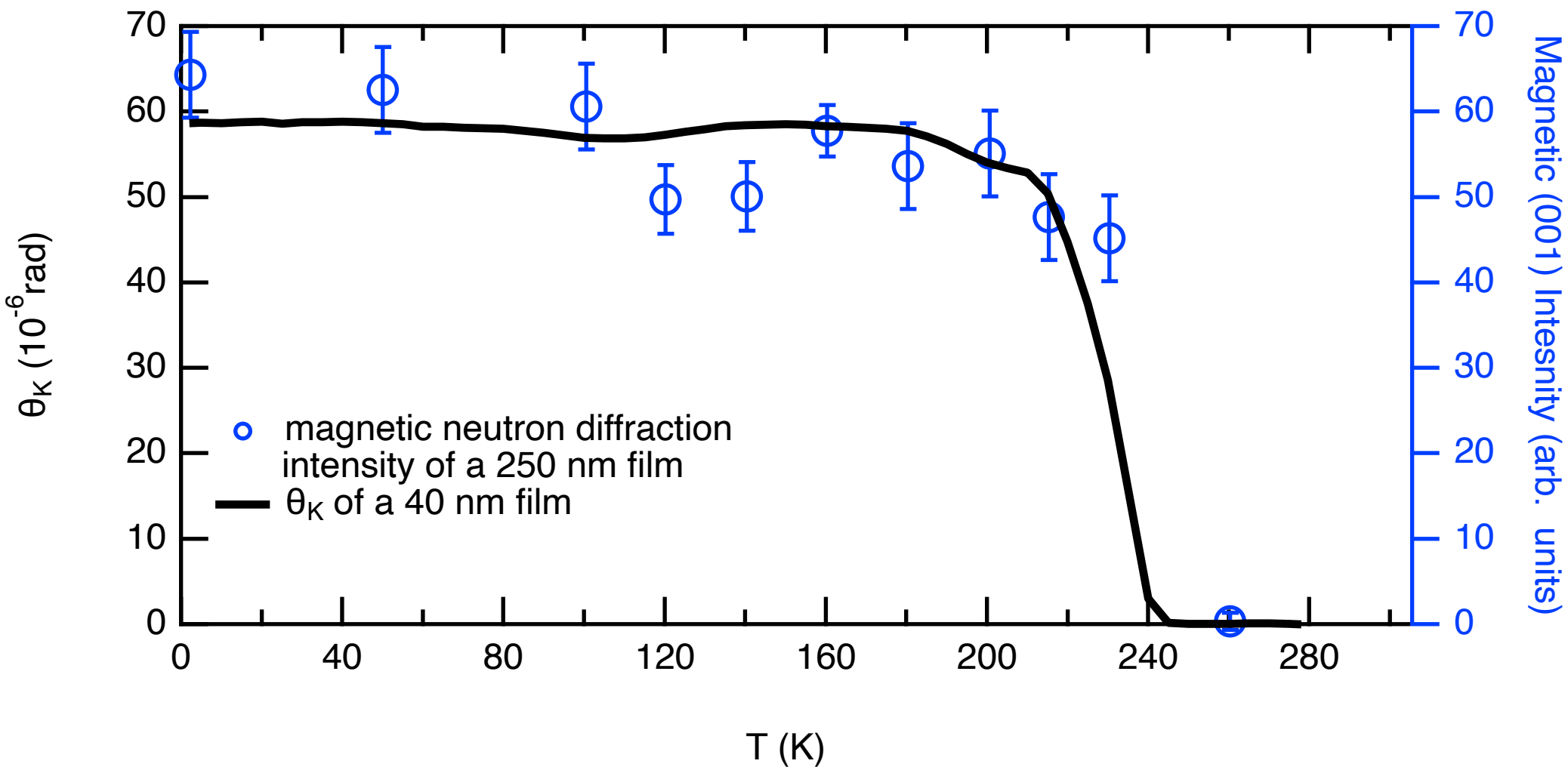

**Figure S1. Magnetic neutron diffraction and MOKE of stoichiometric $Mn_3NiN$ films.** Blue circles are magnetic neutron diffraction intensity of a stoichiometric 250 nm $Mn_3NiN$ film (replotted from Ref [1]). The black line represents the spontaneous MOKE signal $\theta_K$ of a stoichiometric 40 nm thick film, measured during zero-field warm (ZFW) after a *0.3 T* field cool (FC). Both exhibit temperature invariance below the *Neel* temperature.

**Nitrogen deficient film:**

The results reported in the main text are from single crystal epitaxial stoichiometric 40 nm $Mn_3NiN$ films. They are grown on (001)-oriented $(La_{0.3}Sr_{0.7})(Al_{0.65}Ta_{0.35})O_3$ (LSAT) substrates at 650°C under an $Ar/N_2$ (78%:22%) mixed atmosphere of 10 mTorr by DC reactive magnetron sputtering using a stoichiometric $Mn_3Ni$ target.

Bulk crystals of $Mn_3AN$ (A = Ni, Zn, Ga, etc.) are known to be frequently deficient in the interstitial element N [2]. We found that $Mn_3NiN$ thin films grown under low nitrogen partial pressure (<20%) also lose stoichiometry and become nitrogen-deficient. Fig. S2 shows the MOKE signal $\theta_K$ measured in a 40 nm nitrogen-deficient $Mn_3NiN_{1-\delta}$ film during zero-field warming (ZFW) after a 0.3 T field-cooling (FC). While the peak Kerr rotation, $\theta_k \approx 59$ µrad, is comparable to that of the stoichiometric film reported in the main text, the nitrogen-deficient film exhibits a much stronger temperature dependence below 200 *K* and a pronounced peak between 200 *K* and its slightly higher Néel temperature of 255 *K*. We speculate that below 200 *K*, a mixed $\Gamma_{5g}$ phase emerges in nitrogen-deficient films, stabilized by local lattice distortions of the $Mn_6N$ octahedra [3] caused by nitrogen deficiency [2]. Because the $\Gamma_{5g}$ phase carries no Berry curvature [4–6] and thus no Kerr response, the Kerr signal is suppressed until the temperature rises above ~230 *K*, where the $\Gamma_{4g}$ phase once again dominates.

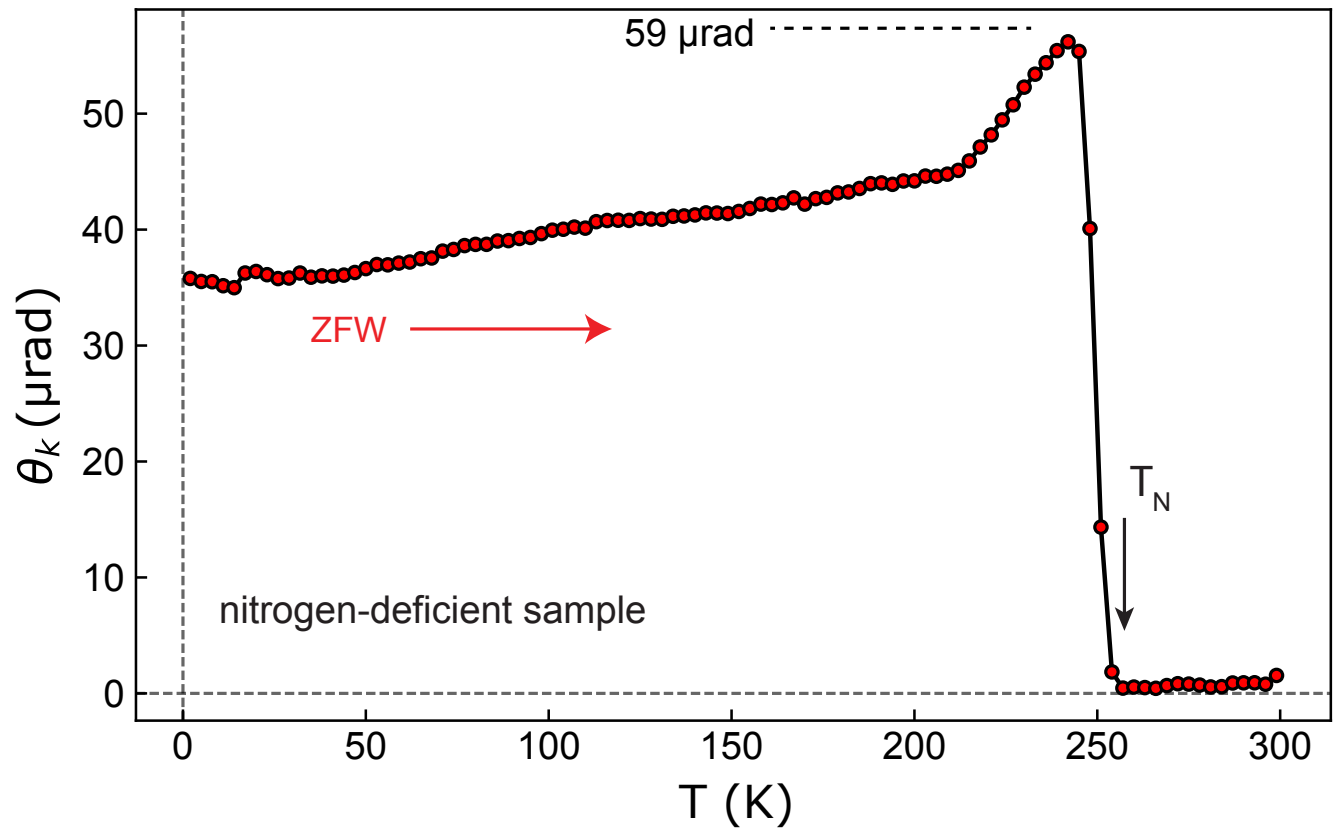

**Figure S2. MOKE of Nitrogen-deficient film $Mn_3NiN_{1-\delta}$.** Shown in red is the spontaneous MOKE signal $\theta_K$ measured in a 40 nm thick nitrogen deficient film during zero-field warm (ZFW) after a *0.3 T* field cool (FC). While the peak $\theta_K$ of 59 $\mu rad$ is comparable to that of the stoichiometric film, it has a much larger temperature variance likely due to a mixed magnetic phase.

It is important to emphasize that all films, whether stoichiometric or nitrogen-deficient, exhibit excellent crystallinity, independent of the nitrogen partial pressure during growth. This is evidenced by the pronounced thickness fringes in Fig.S3a. The stoichiometric film discussed in the main text is with 22% N during growth, while the nitrogen-deficient film in Fig.S2 is with 17% N, which alters the N-composition but keeps Ni and Mn compositions intact. In Fig.S3d, Energy-dispersive X-ray spectroscopy (EDX) confirms that the Mn and Ni compositions are indeed identical in both stoichiometric and nitrogen-deficient films. Due to the inherent limitations of EDX that N is very close to both O and C, it is not possible to determine the nitrogen content in thin films.

Fig.S3b shows the lattice parameters $a_0$ (in-plane) and $c_0$ (out-of-plane) measured by synchrotron XRD. Because the film is under tensile strain from lattice mismatch with the LSAT substrate, $a_0$ is slightly larger than $c_0$. The average lattice ratio $\sqrt[3]{a_0^2\, c_0}$ increases systematically with nitrogen partial pressure, consistent with observations in polycrystalline $Mn_3NiN_{1-\delta}$, where the lattice expands with increasing nitrogen content $1\text{-}\delta$ [2]. Furthermore, due to tensile strain, the lattice volume of stoichiometric films is expected to be slightly larger than the bulk value by Poisson's ratio, which is precisely what we observe in Fig.S3c when the nitrogen partial pressure exceeds 20%.

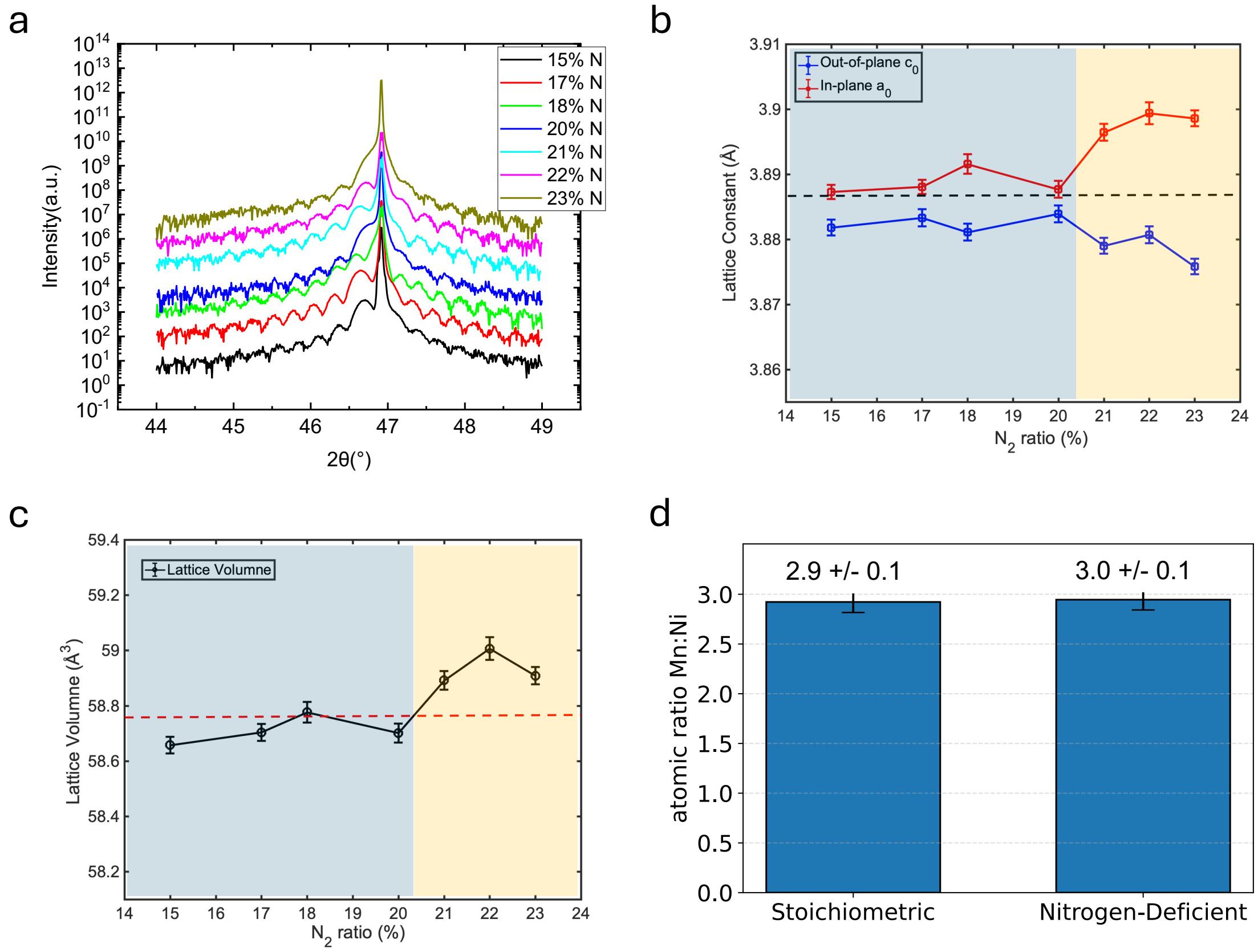

**Figure S3. Structure parameters of stoichiometric and nitrogen-deficient films. a,** Out-of-plane X-ray rocking curves of $Mn_3NiN$ indicating high crystallinity regardless of $N_2$-ratio during growth. The stoichiometric film discussed in the main text is with 22% N during growth, while the nitrogen-deficient film in Fig.S2 is with 17% N. **b, c,** Synchrotron measurements of lattice parameter $a_0$, $c_0$, and lattice volume. Dashed lines represent bulk values. Growth with low $N_2$ partial pressure (< 20%) leads to nitrogen deficiency in the film. **d,** atomic ratio between Mn and Ni determined by Energy-dispersive X-ray spectroscopy (EDX) in the stoichiometric and nitrogen-deficient films.

**Details of AHE measurement:**

Sample's raw longitudinal and transverse resistivities were measured using a resistance bridge. Electric contacts were made with 25 $\mu m$-diameter gold wires that are connected to the samples with silver epoxy. Due to unavoidable deviation from perfect contact alignments, a small percentage of the longitudinal voltage is mixed into the much smaller transverse signal. Conventionally the true Hall resistivity $\rho_{xy}$ is obtained by taking magnetic hysteresis to "symmetrize" the Hall signal, which is not possible in our $Mn_3NiN$ thin films since our maximum field of 9 $T$ becomes insufficient to finish the hysteresis below 175 $K$.

Instead, we use $+0.3\ T$ and $-0.3\ T$ field cooling (FC) from above $T_N$ to polarize the chirality, followed by zero-field warming (ZFW). The difference between raw Hall data $\rho_{xy}(+)$ and $\rho_{xy}(-)$ (Fig. S4a) was taken to remove the mixed time-reversal-even $\rho_{xx}$ component. The resulting Hall resistivity $\rho_{xy} = [\rho_{xy}(+) - \rho_{xy}(-)]/2$ during 0.3 $T$ FC is shown in Fig. S4b. The resistivity matrix is then inverted to obtain the conductivity matrix, namely $\sigma_{xx} = \frac{\rho_{xx}}{\rho_{xx}^2+\rho_{xy}^2}$, and $\sigma_{xy} = -\frac{\rho_{xy}}{\rho_{xx}^2+\rho_{xy}^2}$ .

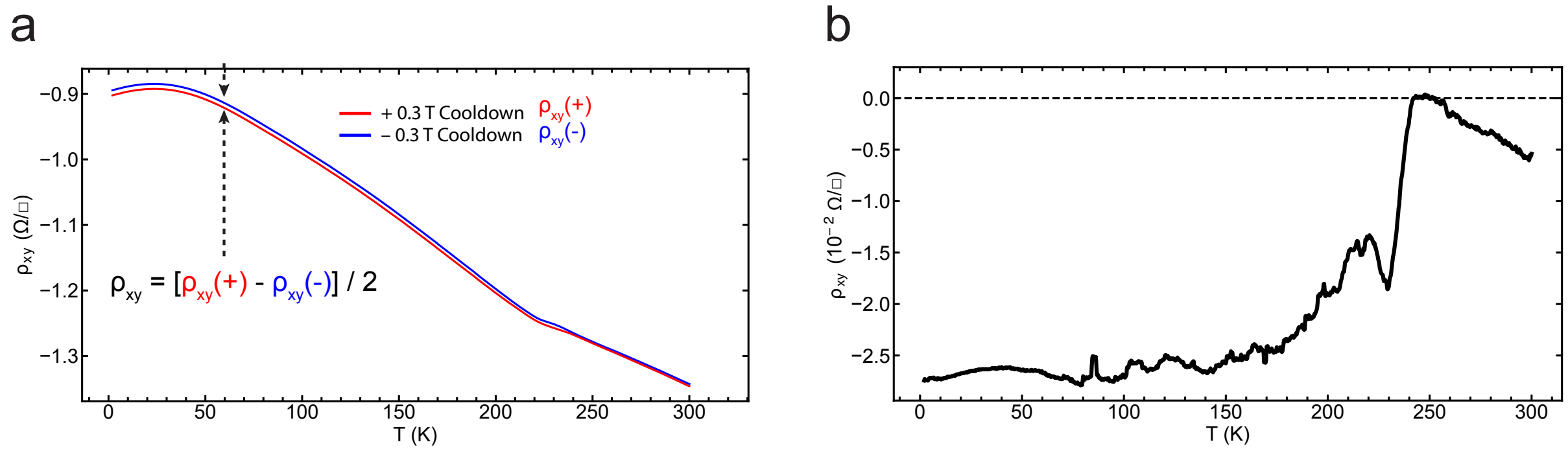

**Figure S4. AHE Measurement. a,** Raw data of Hall resistivities $\rho_{xy}(+)$ (red) and $\rho_{xy}(-)$ (blue) measured during $\pm 0.3\ T$ field cools (FC) that are dominated by the mixed $\rho_{xx}$ component. Their difference yields the true $\rho_{xy} = [\rho_{xy}(+) - \rho_{xy}(-)]/2$. **b,** Obtained $\rho_{xy}$ during 0.3 $T$ FC (black).

**Film Growth:**

Stoichiometric single crystal epitaxial 40 nm $Mn_3NiN$ thin films were grown on (001)-oriented $(La_{0.3}Sr_{0.7})(Al_{0.65}Ta_{0.35})O_3$ (LSAT) substrates at 650°C under an $Ar/N_2$ (78%:22%) mixed atmosphere of 10 mTorr by DC reactive magnetron sputtering using a stoichiometric $Mn_3Ni$ target, which was monitored by in-situ reflection high energy electron diffraction (RHEED).

**Sagnac MOKE measurements:**

MOKE measurements are performed using a zero-loop fiber-optic Sagnac interferometer [7] microscope [8,9]. Upon reflection from the sample the nonreciprocal phase shift $\Delta\varphi$ between the two counterpropagating circularly polarized beams is twice the Kerr rotation $\Delta\varphi = 2\theta_K$. In a zero-loop fiber-optic Sagnac interferometer as shown in Fig.1c. The beam of light is routed by a fiber circulator to a fiber polarizer. After the polarizer the polarization of the beam is at 45° to the axis of a fiber-coupled electro-optic modulator (EOM), which generates 4.6 MHz time-varying phase shifts $\phi_m \sin(\omega t)$, where the amplitude $\phi_m = 0.92$ rad between the two orthogonal polarizations that are then launched into the fast and slow axes of a polarization maintaining (PM) single-mode fiber. Upon exiting the fiber, the two orthogonally polarized linearly polarized beams are converted into right- and left-circularly polarizations by a quarter-wave plate (QWP) and are then focused onto the sample. After reflection from the sample, the same QWP converts the reflected beams back into linear polarizations with exchanged polarization axes. The two beams then pass through the PM fiber and EOM but with exchanged polarization modes in the fiber and the EOM. At this point, the two beams have gone through the same path but in opposite directions, except for a phase difference of $\Delta\varphi$ from reflection off the magnetic sample and another time-varying phase difference by the modulation of EOM. This nonreciprocal phase shift $\Delta\varphi$ between the two counterpropagating circularly polarized beams upon reflection from the sample is twice the Kerr rotation $\Delta\varphi = 2\theta_K$. The two beams are once again combined at the detector and interfere to produce an optical signal $P(t)$:

$$P(t) = \frac{1}{2} P[1 + \cos(\Delta\varphi + \phi_m \sin(\omega t))] \tag{1}$$

, where $P$ is the returned power if the modulation by the EOM is turned off. For MOKE signals that are slower that the 4.6 MHz modulation frequency used in this experiment, we can treat $\Delta\varphi$ as a slowly time-varying quantity. And $P(t)$ can be further expanded into Fourier series with the first few orders listed below:

$$\begin{aligned} P(t)/P = \; & \frac{1}{2}\,[1 + \mathrm{J}_0(2\phi_m)] \\ & +(\sin(\Delta\varphi)\, J_1(2\phi_m))\sin(\omega t) \\ & +(\cos(\Delta\varphi)\, J_2(2\phi_m))\cos(2\omega t) \\ & +2\,\mathrm{J}_3(2\phi_m))\sin(3\omega t) \\ & +\cdots \end{aligned} \tag{2}$$

, where $\mathrm{J}_1(2\phi_m)$ and $\mathrm{J}_2(2\phi_m)$ are Bessel J-functions. Lock-in detection was used to measure the first three Fourier components: the average (DC) power ($P_0$), the first harmonics ($P_1$), and the second harmonics ($P_2$). And the Kerr rotation can then be extracted using the following formula:

$$\theta_K = \frac{1}{2}\,\Delta\varphi = \frac{1}{2}\tan^{-1}\left[\frac{\mathrm{J}_2(2\phi_m)P_1}{\mathrm{J}_1(2\phi_m)P_2}\right] \tag{3}$$

**Thermal broadening estimate:**

The temperature dependence of the intrinsic off-diagonal conductivity in Eq. (1) enters only through the Fermi Dirac function f. Writing the zero-temperature anomalous Hall conductivity as a function of the Fermi energy, $\tilde{\sigma}_{xy}(\varepsilon)$, the finite-temperature conductivity is the thermally weighted average:

$$\sigma_{xy}(T) = \int d\varepsilon\, (-\partial f/\partial\varepsilon)\, \tilde{\sigma}_{xy}(\varepsilon),$$

which, by a Sommerfeld expansion, gives the leading thermal-broadening correction of

$$\sigma_{xy}(T) \approx \tilde{\sigma}_{xy}(\mu) + (\pi^2/6)(k_B T)^2\, \tilde{\sigma}_{xy}''(\mu)$$

The fractional thermal-broadening contribution is therefore

$$\delta \approx (\pi^2/6)(k_B T)^2\, |\tilde{\sigma}_{xy}''(\mu)/\tilde{\sigma}_{xy}(\mu)| \sim (\pi^2/6)(k_B T/W)^2,$$

where W is the energy scale over which $\tilde{\sigma}_{xy}(\varepsilon)$ varies near the Fermi level, set by the band and spin orbit features (anticrossings and small gaps) that host the Berry-curvature hot spots. The thermal energy is $k_B T$ ~ 0.17 meV at 2 K, ~ 8.6 meV at 100 K, and ~ 18 meV at 210 K. For $Mn_3NiN$, with the relevant band/spin–orbit features on the scale W ~ 0.2 to 0.5 eV, the estimated thermal broadening of the intrinsic conductivity is $\delta < 1\%$ up to 210 K and well below this at low temperature. This is fully consistent with the observed temperature invariance of the spontaneous Kerr signal (varying by less than 1% below 100 K and within a few percent up to 210 K).

In contrast, the measured dc anomalous Hall conductivity increases by roughly a factor of five across the antiferromagnetic phase (from ~5 to ~27 S/cm), a change about two orders of magnitude larger than the thermal-broadening estimate above. Thermal smearing of the intrinsic conductivity therefore cannot account for the strong temperature dependence of the dc signal. Instead, this dependence is dominated by extrinsic skew scattering, as confirmed by the scaling analysis $\rho_{xy} = a\, \rho_{xx} + b\, \rho_{xx}^2$ (main text, Fig. 4c), in which the large linear (skew-scattering) term a accounts for the bulk of the temperature variation while the constant term b captures the temperature-independent intrinsic contribution.

An independent empirical bound is provided by the optical data themselves: Fermi–Dirac broadening affects the diagonal conductivity $\sigma_{xx}$ as well as $\sigma_{xy}$, yet the measured reflectivity and hence $\sigma_{xx}$(0.8 eV) are temperature-independent to within 1% from 1.8 to 300 K (Fig. 3c, e). This directly limits the magnitude of thermal-broadening effects on the optical conductivity to the percent level, corroborating the estimate above without recourse to a band-structure calculation.

We emphasize that this argument concerns the temperature dependence of the conductivity, not its absolute magnitude. The dc $\sigma_{xy}$ (the $\omega \to 0$ Berry-curvature integral) and the optical $\sigma_{xy}$(0.8 eV) (a finite-frequency interband response) sample the conductivity at different frequencies and are not expected to coincide in magnitude. Only their contrasting temperature dependence is at issue here: strong in dc transport, absent in the optical response.